\documentclass[12pt,rotating]{elsart}
\usepackage{rotating}

\usepackage{cites,setspace,url}
\usepackage{epsfig}

\bibnum{k}

\begin{document}

\citenum{}

\begin{frontmatter}

\title{Simplified PMT Model}

\author{Karim Zbiri*}
\corauth{Tel.: +1 215 895 1887; +1 215 895 2990.\\
E-mail address: zbiri@physics.drexel.edu\\
 (K. Zbiri). 
}

\maketitle

\address{*Department of physics, Drexel University, 3141 Chestnut St,
Philadelphia, PA 19104, USA.}

\begin{abstract}
A simplified model is proposed based on the characteristics
 of the photomultiplier tube (PMT). The Model is compared to the 
available data, and  it successes to reproduce the saturation profile of 
the PMT. The model also adds  clarification about how the PMT works.
\\


\end{abstract}

\end{frontmatter}

\section{Introduction}

At Drexel University, I was testing the 8-inch, 13 stages, R1408 PMTs 
which will
 be used for
the inner vetos of the Double Chooz experiment \cite{DCLoi}. These PMTs were 
used
in the IMB \cite{IMB} and SuperKamiokande \cite{superk} experiments.
This testing allowed a better understanding of their working. In this paper I propose a model
inspired from observations during the testing period. The first
aim of this model is to explain in simple way  the 
saturation characteristics of the PMT.
The Model gives a global description 
of how the different PMTs characteristics, like the quantum
efficiency (QE) or the collectivity efficiency (CE),
are linked together to the PMT gain (G). The model is compared to the data
obtained from three different PMTs, which have different single
photoelectron (spe) distributions.

\section{The general case}

The basic idea is   at a fixed applied high voltage, the PMT works as
 charge amplifier, which will amplify
 below a certain charge limit $q_{sat}$ and beyond this limit the
PMT will continue to amplify the signal but at lower gain. As the
signal in the input of the PMT is increased, the PMT gain decreases.

A general profile which  matches this description of the gain
variation is given by:

\begin{equation}
\label{gain}
G=\frac{G_0}{\sqrt{1+q^\alpha/q^\alpha_{sat}}},
\end{equation}

Where $\alpha$ is a real coefficient which can be  determined from
the comparison with the data, $q$ is the collected charge at
the anode and $G_0$ is the gain at the limit as  $q \rightarrow 0$ , 
and depends only on the applied high voltage like following \cite{HAM06}:

\begin{equation}
\label{G0}
G_0=k V^{\beta N}.
\end{equation}
Where k is a constant, which depends on the material of the dynodes and the
voltage division between them, N is the number of the dynodes
and $\beta$ is between 0.7 and 0.8. $G_0$ is usually set up by appropriate high
voltage at low light level (single photoelectron
level). The $q_{sat}$ is an intrinsic constant to each PMT, and it depends only
on the type of the voltage divider. $q_{sat}$ fixes the
amplification performance of the PMT at a given high voltage, and
it determines the
collectivity efficiency of the PMT. The amplification performance can be
adjusted by the use of the appropriate voltage divider.\\

\section{Comparisons of the model with the measurements}

\subsection{The R1408 PMT case}

\begin{figure}
\begin{center}\includegraphics[%
  scale=0.5]{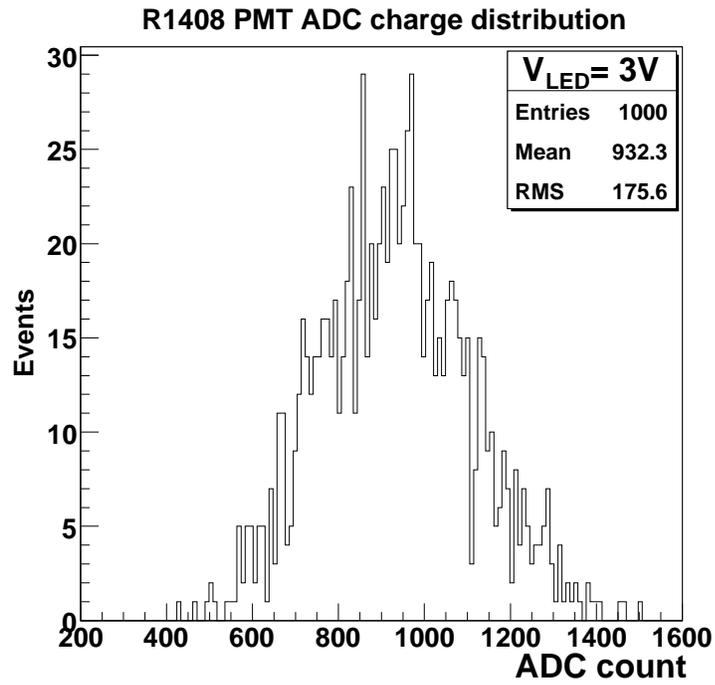}\end{center}

\caption{\label{cap:R1408_LED_pulse_LIN} ADC distribution of the PMT
pulses for LED flashing with 3V from pulser.}
\end{figure}

\begin{figure}
\begin{center}\includegraphics[%
  scale=0.5]{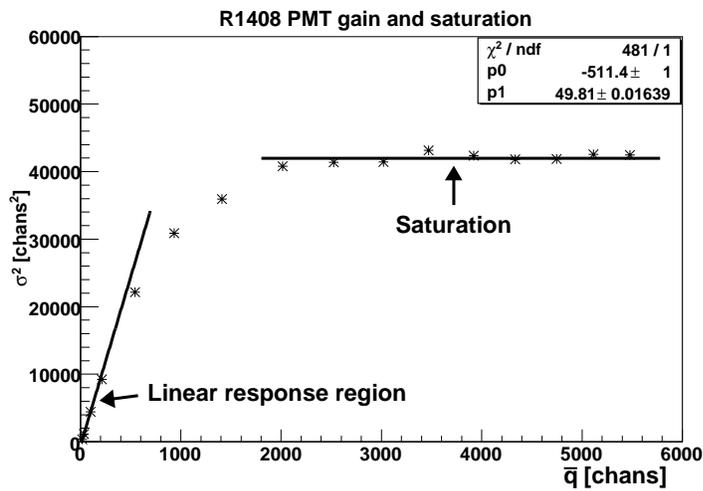}\end{center}

\caption{\label{cap:R1408_LIN_gain} ADC variance($\sigma ^2$) versus ADC
mean($\bar{q}$) for a variety pulse heights. The slope of the fit of the linear
region gives a gain/channel=slope/2=24.9 ~channel/photoelectron.}
\end{figure}

\begin{figure}[h]
\begin{center}\includegraphics[%
 scale=0.5 ]{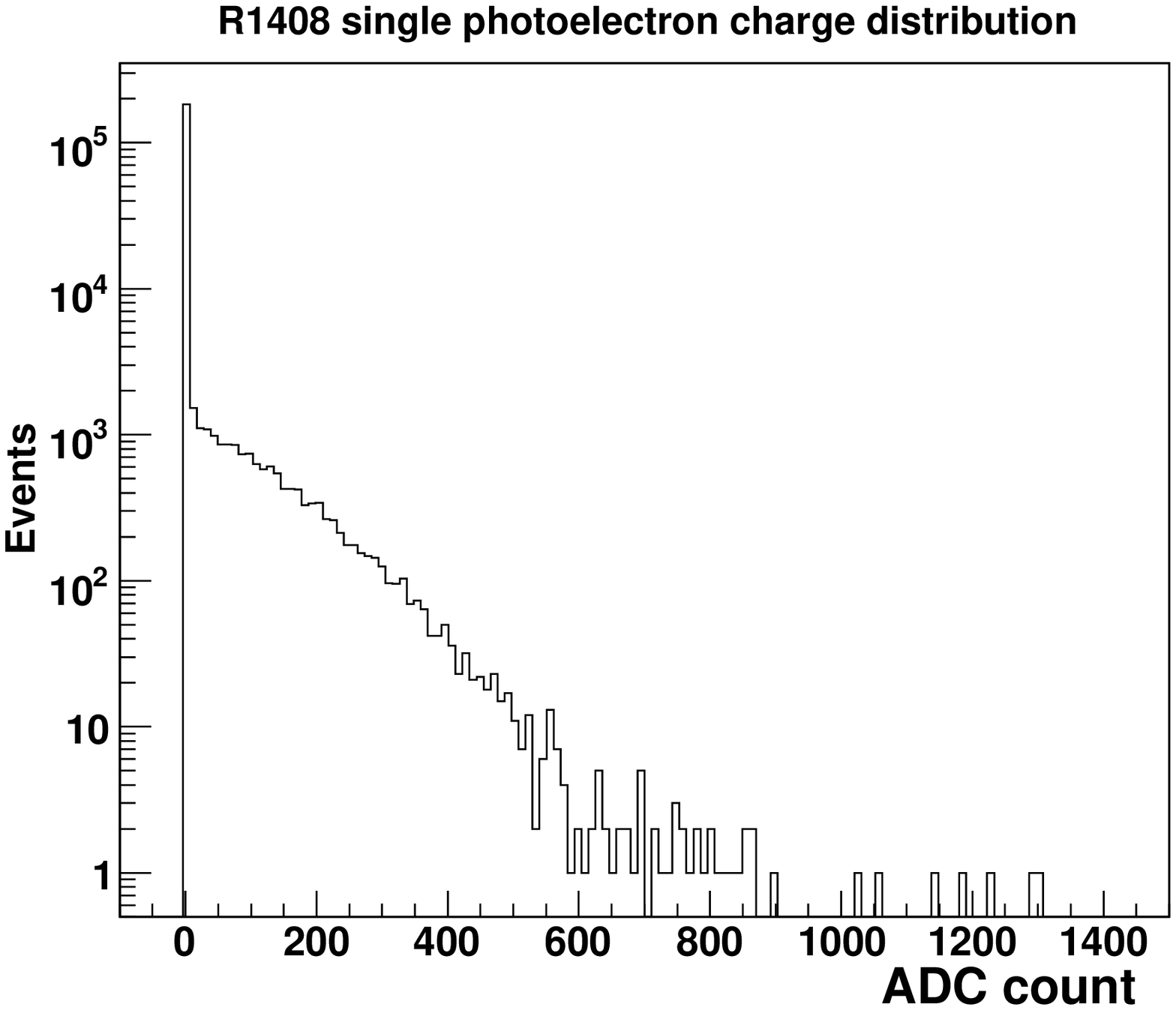}\end{center}
\caption{\label{cap:R1408_spe} The single photoelectron (spe) charge
distribution for the 8-inch, 13 stages, R1408 PMT. The ADC pedestal is not
removed. The signal from the PMT output is amplified 10 times.}
\end{figure}

The 8-inch Hammamatsu R1408 PMT with 13 stages does not have a single
photoelectron peak.
Figure \ref{cap:R1408_spe}
shows the R1408 single
photoelectron (spe) charge distribution.
The main method to determine the gain G  for this
kind of PMTs
is the photo-statistics method \cite{Lane08}. A series of
LED/laser runs are taken at varying light levels, and  for each run 
the statistical mean and variance of the spectra are measured as shown
in Fig. \ref{cap:R1408_LED_pulse_LIN}, after that plot the variance versus
the mean, and fit the linear region with a polynomial of the first order like
in Fig. \ref{cap:R1408_LIN_gain}. The gain is extracted from
the slope of the fit following the equation \cite{Lane08}:
\begin{equation}
\label{PHgain}
\frac{d\sigma^2}{dE}=2eG,
\end{equation}

Where E and $\sigma^2$ are the mean and the variance of the ADC charge
distribution, given by the following relations:

\begin{equation}
\label{expectation}
E=q+\mu_p
\end{equation}
\begin{equation}
\label{variance}
\sigma^2=2qeG+ \sigma^2_p,
\end{equation}

with $e=1.6~10^{-19}C$ is the electron charge, q is the mean charge
at the input
of the ADC.   $\mu_p$ and $\sigma^2_p$ are the mean and
the variance of the ADC pedestal distribution.
One important feature of the photo-statistics method is that 
it is ADC pedestal independent.

\begin{figure}
\begin{center}\includegraphics[%
 angle=-1,scale=1. ]{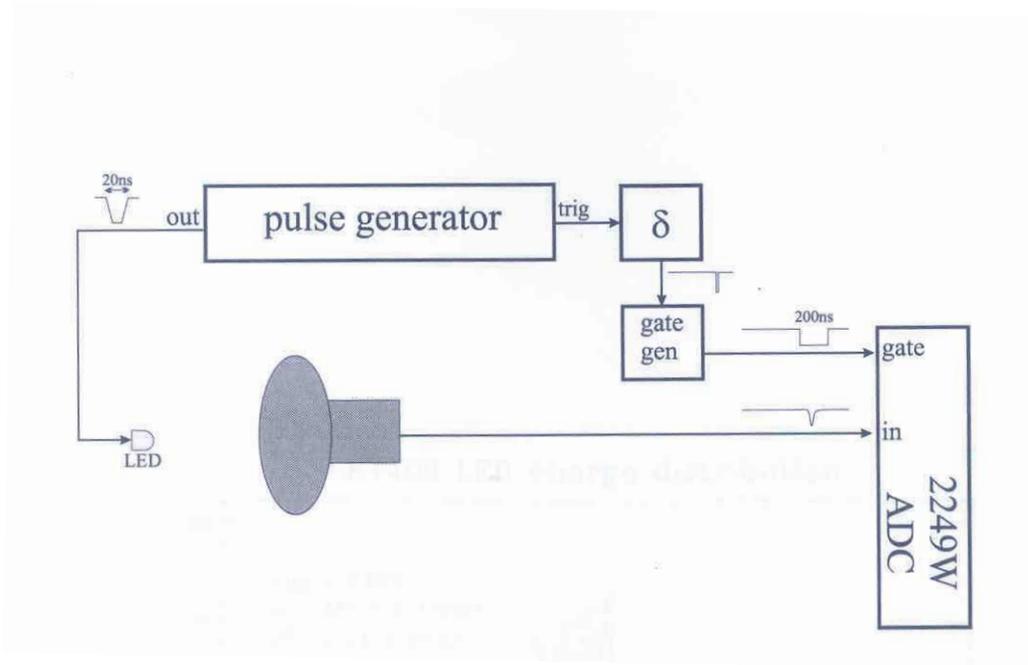}\end{center}
\caption{\label{cap:setup} The test setup scheme.}
\end{figure}

The R1408 PMT was put  under high voltage inside a dark box, illuminated
by  an LED
driven by a fast pulser. The trigger output of the pulser was sent to
Lecroy 222 Gate and Delay units, first to set a delay, and then to
produce a 200 ns gate which enclosed the PMT pulse that occurred when
the LED was flashed, the using setup is shown in Fig. \ref{cap:setup}.
The PMT signal was sent to the input   of a Lecroy 2249W ADC,
while the gate was coming from the 200 ns gate and delay output. 
The LED was flashed
at 20 Hz at a variety of driving voltages from 1V to 8V, all with very narrow
(20 ns) driving pulse widths. In all cases, the PMT pulse was observed on an
oscilloscope to be completely contained within the 200ns ADC gate. For each
of the ADC distributions, the mean and variance were calculated, as it 
discussed above, and the ADC variances were plotted as a function of
the ADC means which allowed determination of the gain $G_0$. The value
of $G_0$ was calculated from the linear region obtained at low LED
driving voltages.

To compare the model to the data,  the variance need to be written  as function
of the measured charge q. If the equation \ref{variance} is
combined with the equation \ref{gain}, we obtain the following relation:

\begin{equation}
\label{generalVar}
\sigma^2= 2eG_0\frac{q} {\sqrt{1+\frac{q^{\alpha}}{q_{sat}^{\alpha}}}}.
\end{equation}

Because the equation \ref{generalVar} is the theoretic variance, it does 
not contain any ADC pedestal variance term.

For large q, the variance can be written like following:
\begin{equation}
\label{InfinityVar}
\sigma^2_{\infty}=2 e G_0 q_{sat}^{\frac{\alpha}{2}} q^{1-\frac{\alpha}{2}}.
\end{equation}

There are three cases of interest:\\
 \begin{enumerate}
 \item $\alpha < 2: \sigma^2_{\infty} \rightarrow \infty. $
 \item $\alpha = 2: \sigma^2_{\infty} \rightarrow 2eG_0q_{sat}.$
 \item $\alpha > 2: \sigma^2_{\infty} \rightarrow 0. $
 \end{enumerate}

\begin{figure}[h]
\begin{center}\includegraphics[%
 scale=0.5 ]{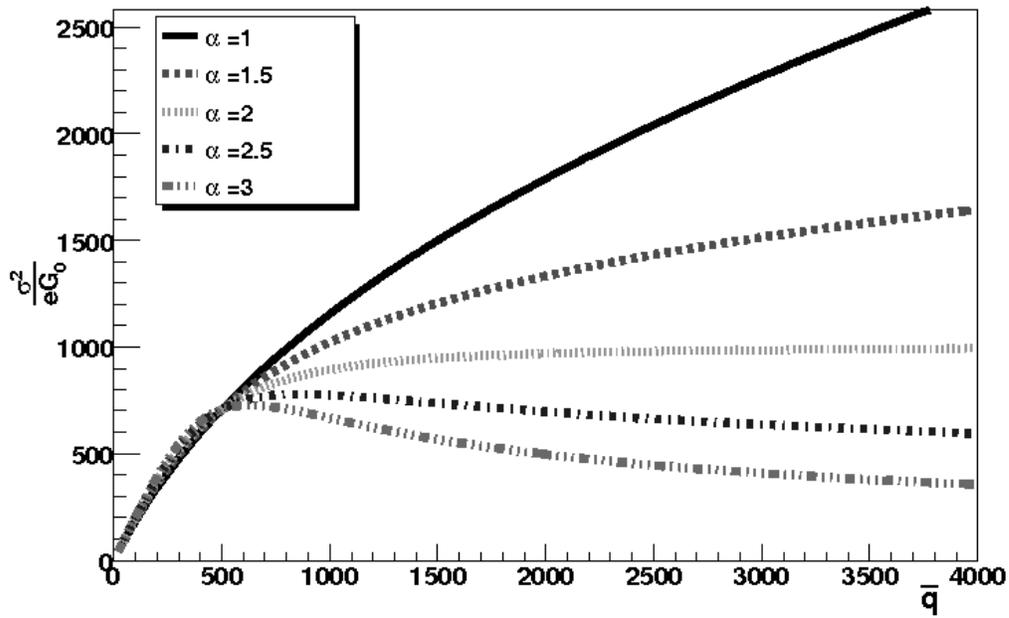}\end{center}
\caption{\label{cap:Variance_q} The ratio $\frac{\sigma^2}{eG_{0}}$
as function of the  mean of the collected charge  at the anode $\bar{q}$, 
as given by the equation
 \ref{generalVar} with $q_{sat}=500$.}
\end{figure}

Figure \ref{cap:Variance_q} shows the variation of the variance
obtained from equation
\ref{generalVar} as function of q, for different values of $\alpha$.\\

\begin{figure}
\begin{center}\includegraphics[%
 scale=0.43 ]{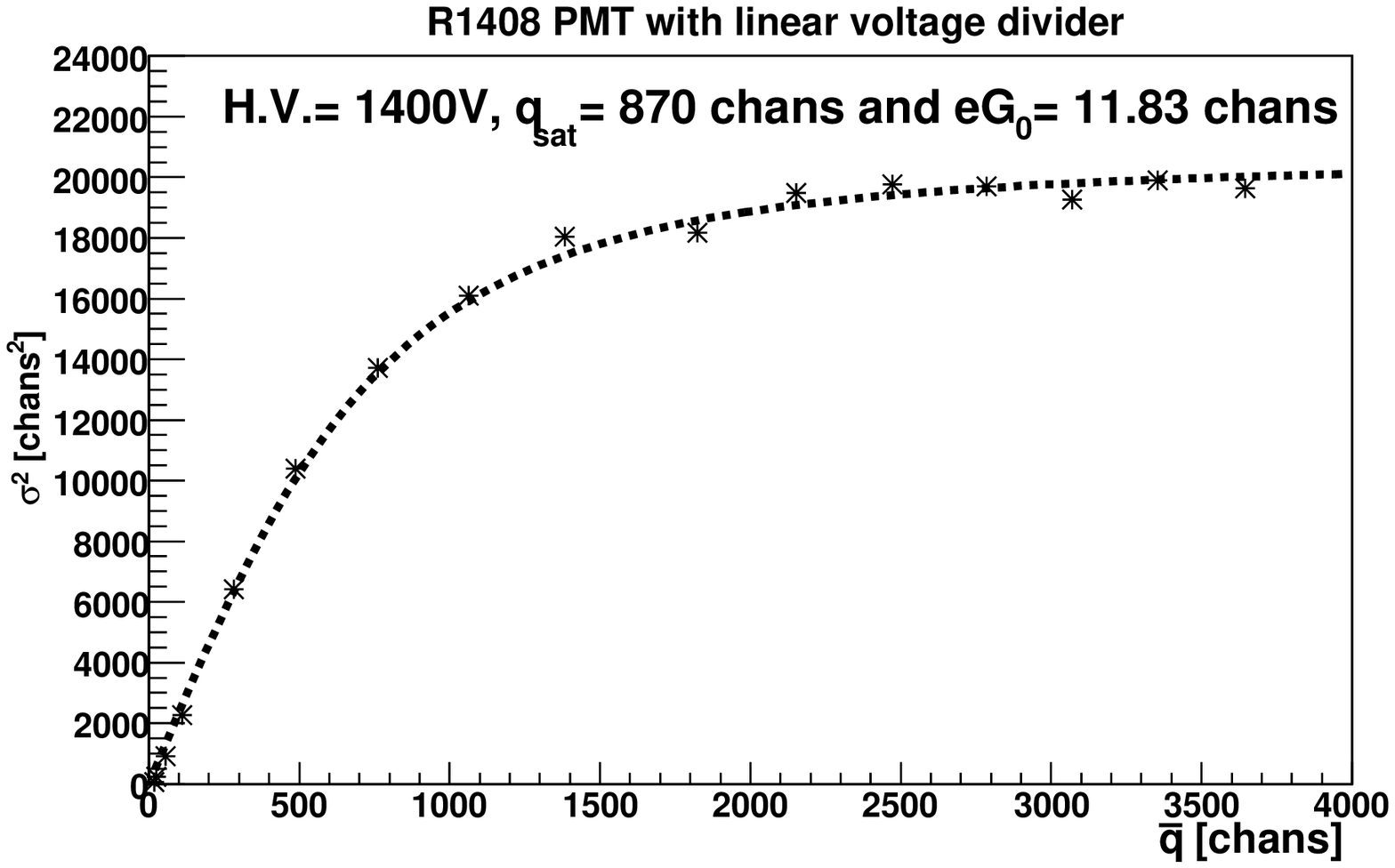}\end{center}
\begin{center}\includegraphics[%
 scale=0.43 ]{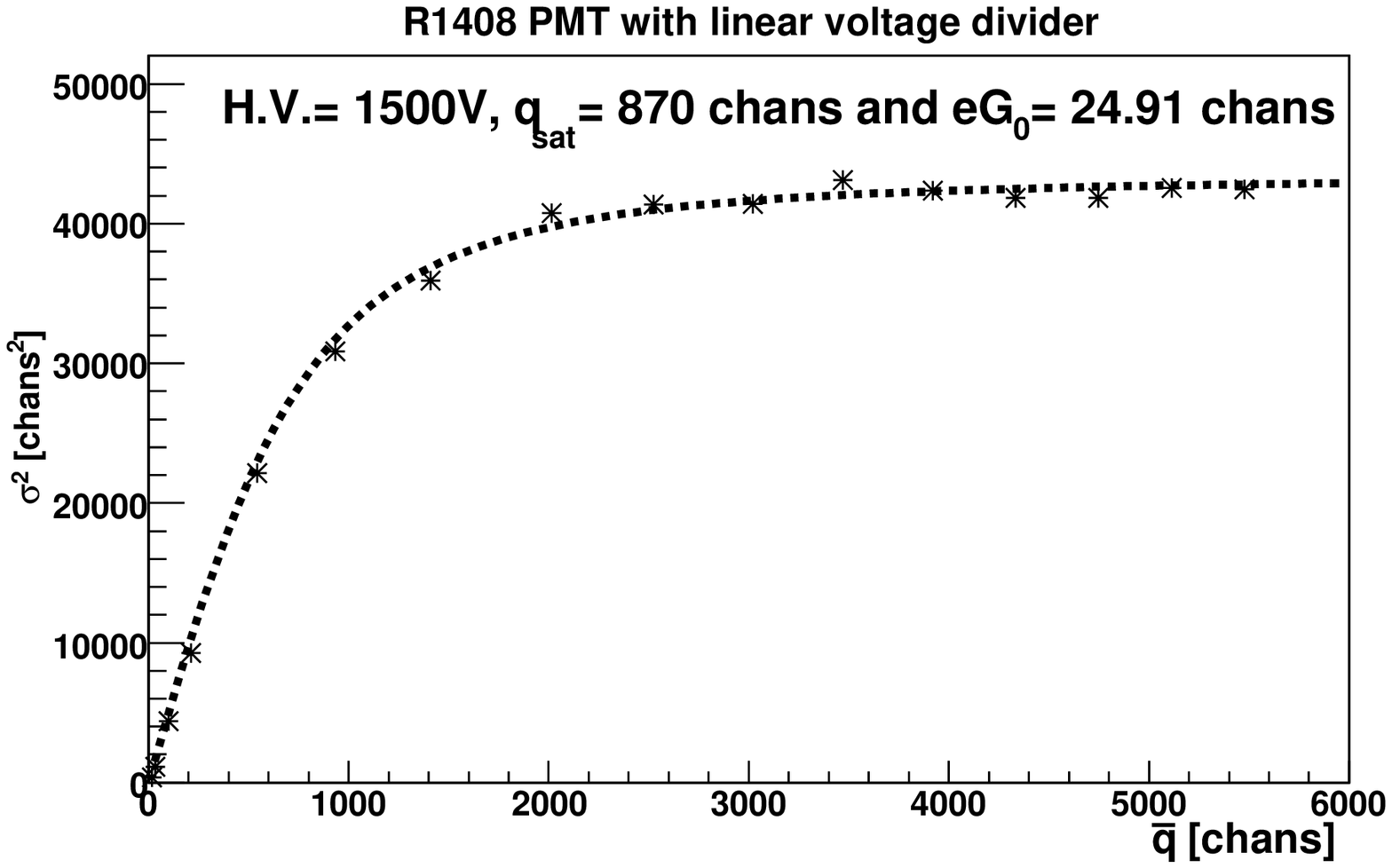}\end{center}
\begin{center}\includegraphics[%
 scale=0.43 ]{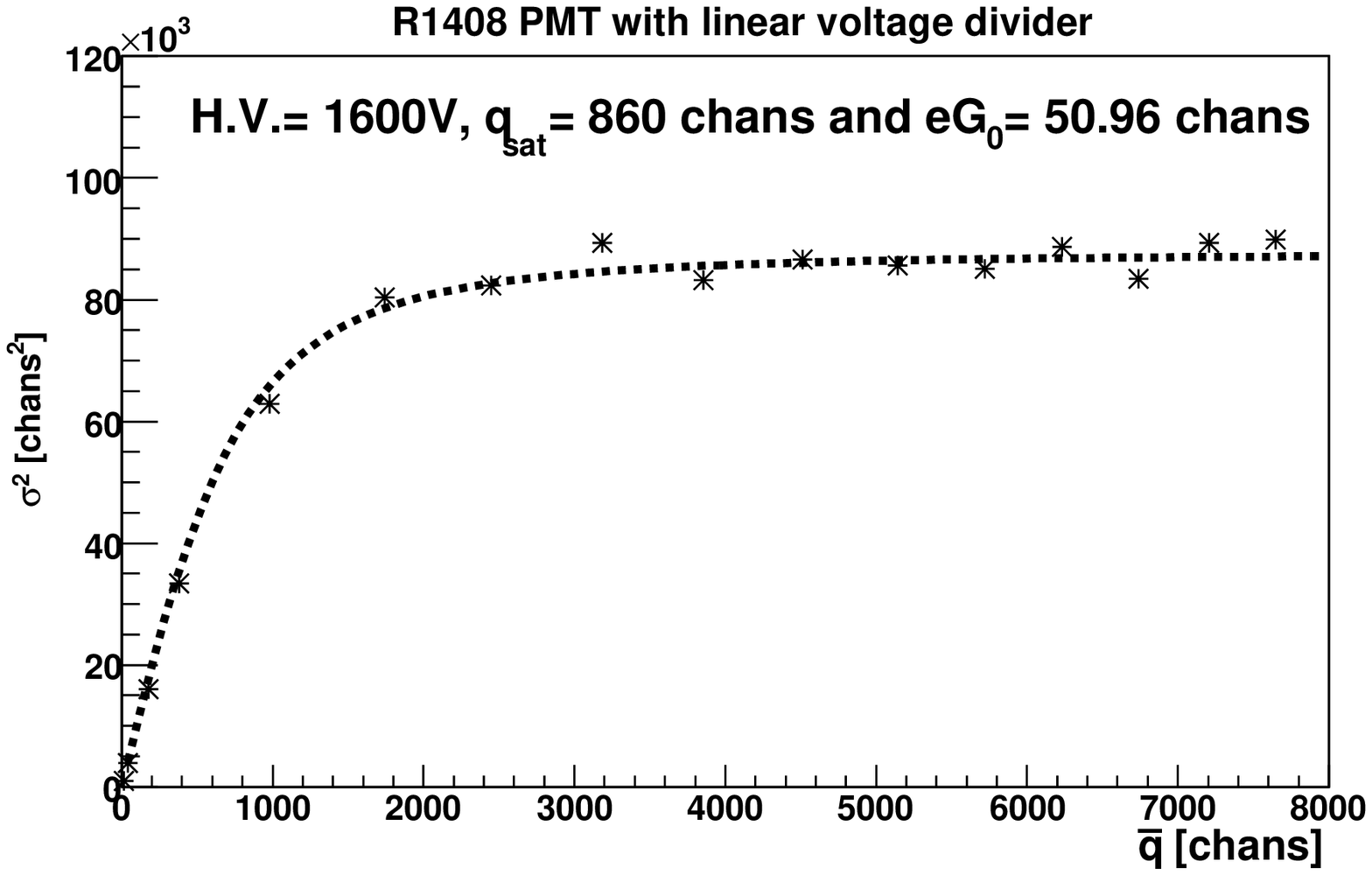}\end{center}
\caption{\label{cap:R1408_LIN_1600V} Comparison of the model with the data
obtained at different high voltages, $q_{sat}$ and $eG_0$ are in ADC channels.
The stars represent the data, and the dashed line represents the model with
$\alpha= 2$ for all.}
\end{figure}

As shown in the Fig. \ref{cap:R1408_LIN_gain} the variance shows a kind of
saturation for high collected charge $q$, which corresponds to the case
$\alpha = 2$.\\
Figure \ref{cap:R1408_LIN_1600V} shows
the comparison of the model with the data obtained from the R1408 PMT at
different applied high voltages,
this PMT has a linear (or equally distributed) voltage divider. The
reference \cite{HAM06} has more details about the 
voltage dividers and their different types, especially chapter 5.
The value of $G_0$ depends on the applied high voltage,
and it is determined as explained above from the slope of the linear region.
The value of $q_{sat}$ is adjusted to reproduce the data, and
the model reproduces the measurements with $\alpha =2$.

\subsection{Effect of the high voltage divider on the saturation level}
\begin{figure}
\begin{center}\includegraphics[%
 scale=0.43 ]{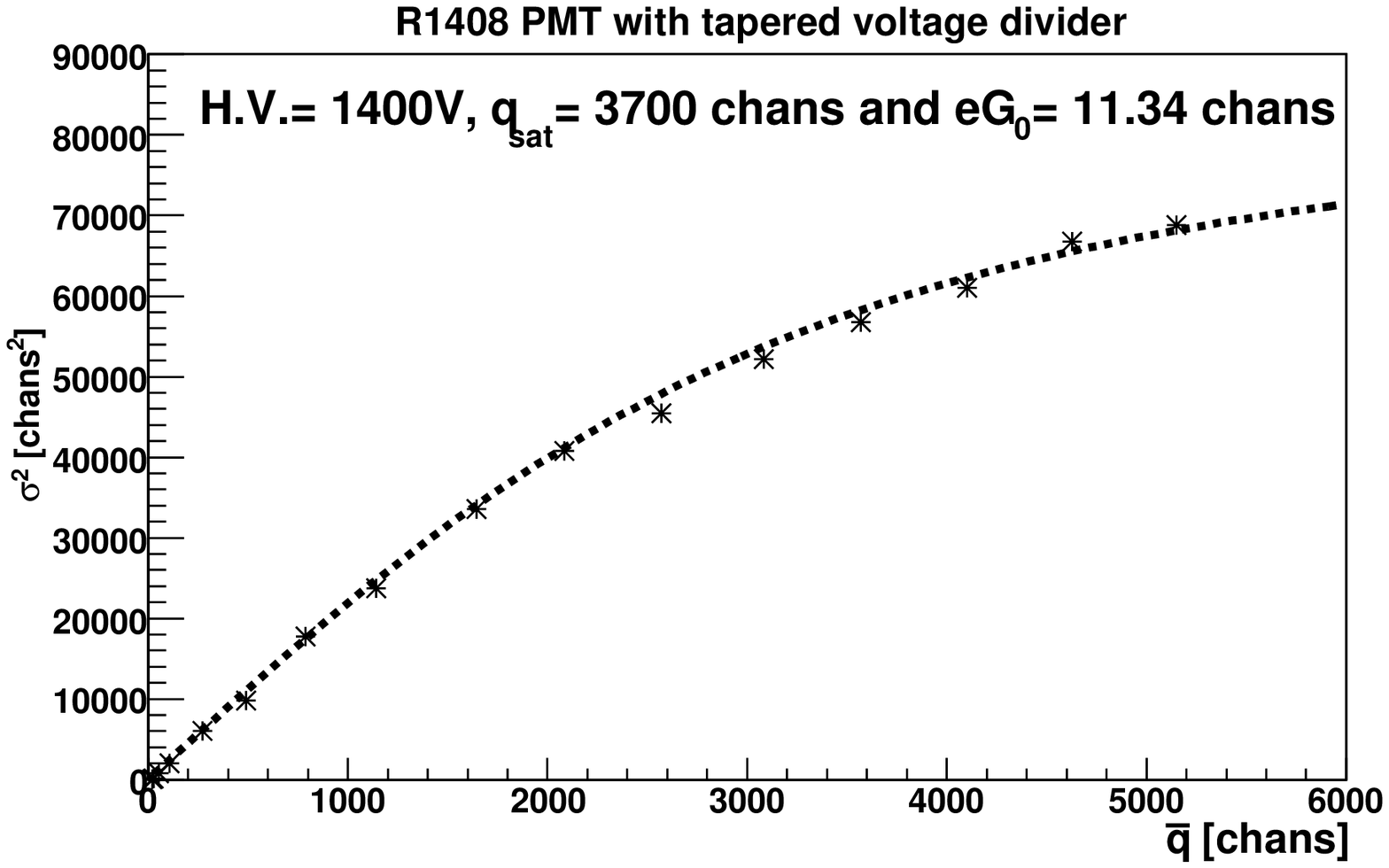}\end{center}
\begin{center}\includegraphics[%
 scale=0.43 ]{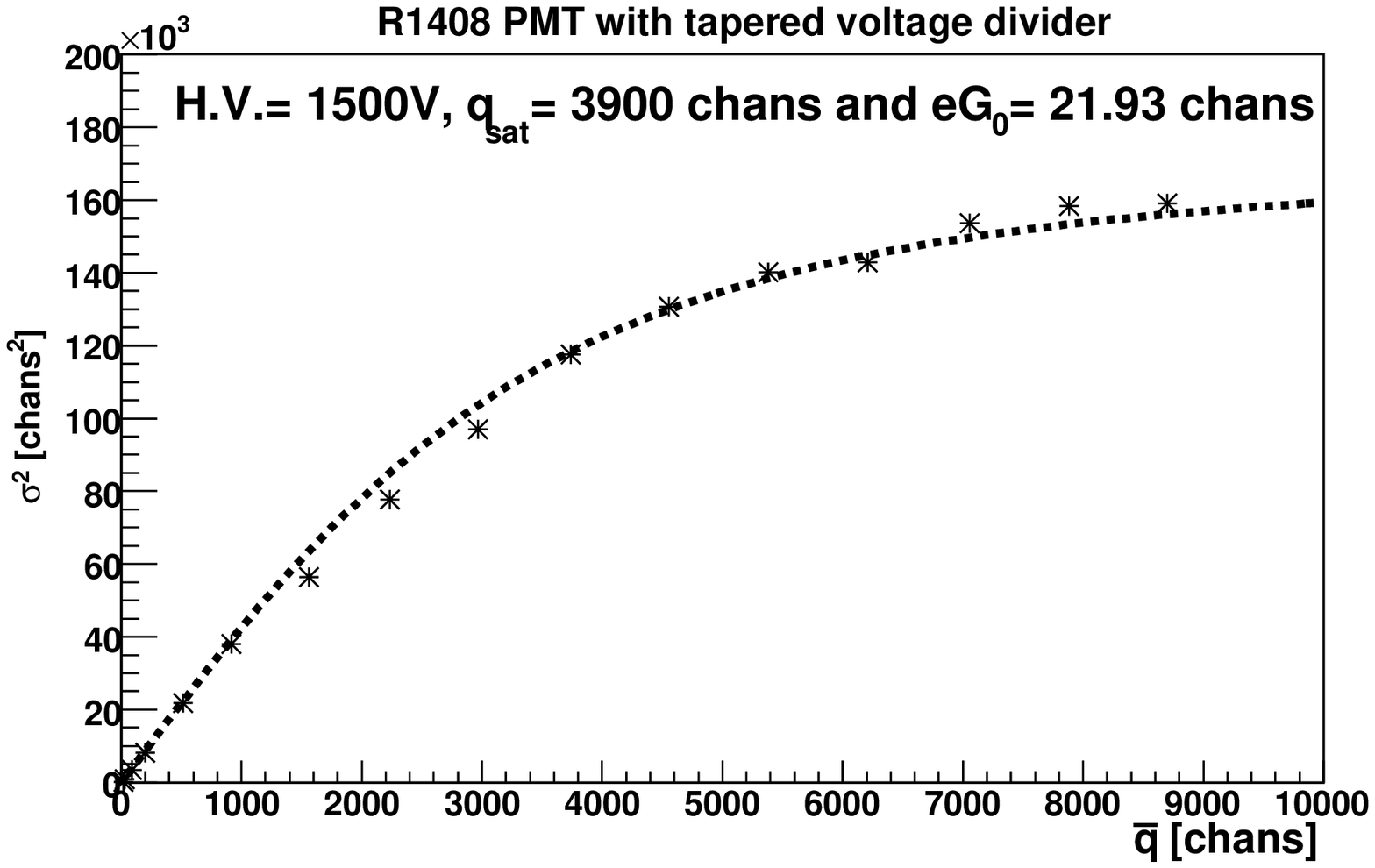}\end{center}
\begin{center}\includegraphics[%
 scale=0.43 ]{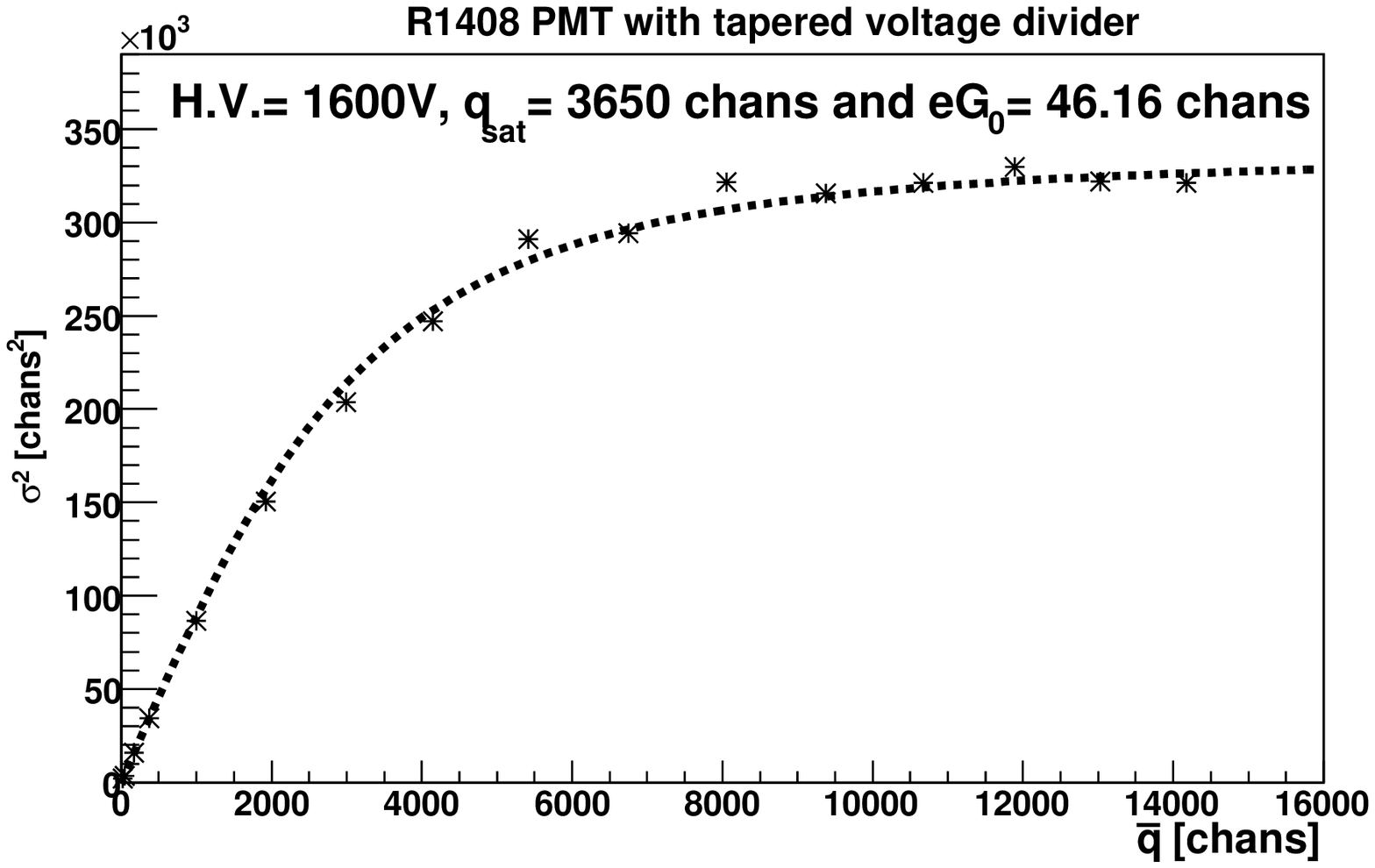}\end{center}
\caption{\label{cap:R1408_YK_1600V} Comparison of the model with the data
obtained at different high voltages, $q_{sat}$ and $eG_0$ are in ADC channels.
The stars represent the data, and the dashed line represents the model with
$\alpha= 2$ for all.}
\end{figure}

To improve the saturation performance of the PMT, a tapered
voltage divider \cite{HAM06} is used instead of the linear one,  
 but the gain is expected to become lower
 for the same total applied high voltage.
Details about the linear and tapered voltage divider, which were 
used during the
measurements, will be given in a separate paper about the Double Chooz inner
veto PMTs testing performed at Drexel University \cite{Lane08}.
Figure \ref{cap:R1408_YK_1600V} shows the
comparison of the model with the data obtained from the measurements with
the R1408 PMT at different applied high voltages, this time the PMT has a
tapered voltage divider. As expected the tapered voltage divider increases the
value of $q_{sat}$, which became at least 4 times larger. In addition, 
The model reproduces the measurements with $\alpha =2$. 

I want conclude this section with one remark about the effect of the
voltage divider. Since the tapered divider enhances the collectivity
efficiency, the gain can be expected to become lower at the same applied 
high voltage the gain due to the
increasing of the collected charge at the anode. The assumption will be
true if the
gain depends only on the collected charge at the anode. However as it 
proposed by
the equation \ref{gain}, the gain does not depend on the collected charge q,
but it  depends on the ratio $\frac{q}{q_{sat}}$. The tapered voltage
divider will increase q and $q_{sat}$ at the same time, then if they are
increased in the same way, their ratio will remain constant
and the gain will not be different from the one obtained with the linear
voltage divider at the same applied total high voltage.

\subsection{The R5912 PMT case}
\begin{figure}[h]
\begin{center}\includegraphics[%
 scale=0.5 ]{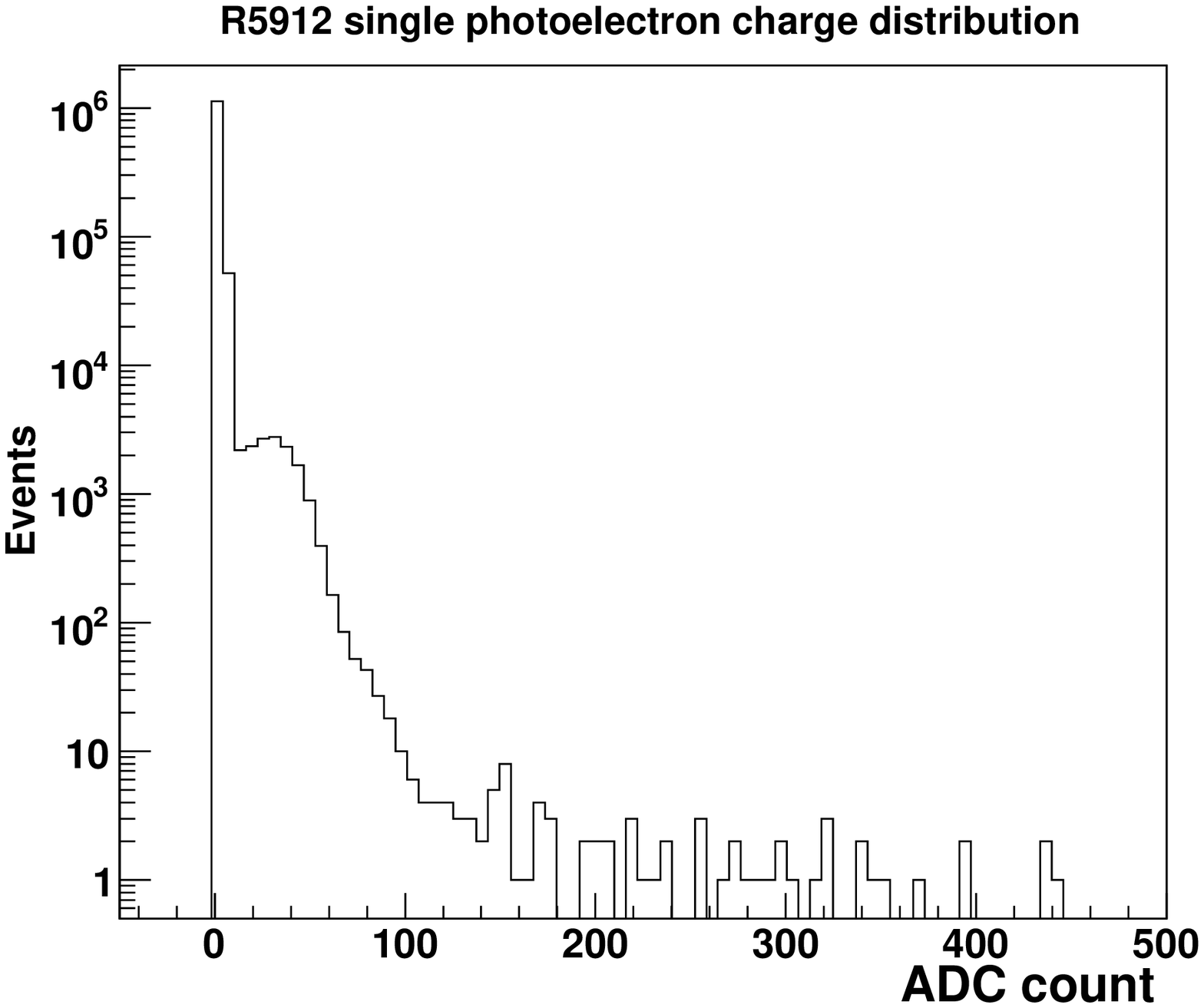}\end{center}
\caption{\label{cap:R5912_spe} The single photoelectron (spe) charge
distribution for the 8-inch R5912 PMT. The ADC pedestal is not removed.
The signal from the PMT output is amplified 10 times.}
\end{figure}

The 8-inch Hammamatsu R5912 PMT has a single photoelectron peak as 
shown in the Fig. \ref{cap:R5912_spe}.
In order to apply the model to the case, a general formulation of the 
variance has to be found.
The equation \ref{variance} can be written
in a general way like following \cite{Lane08}:

\begin{equation}
\label{Genvar}
\sigma^2=\gamma qeG+ \sigma^2_p.
\end{equation}

\begin{figure}
\begin{center}\includegraphics[%
 scale=0.43 ]{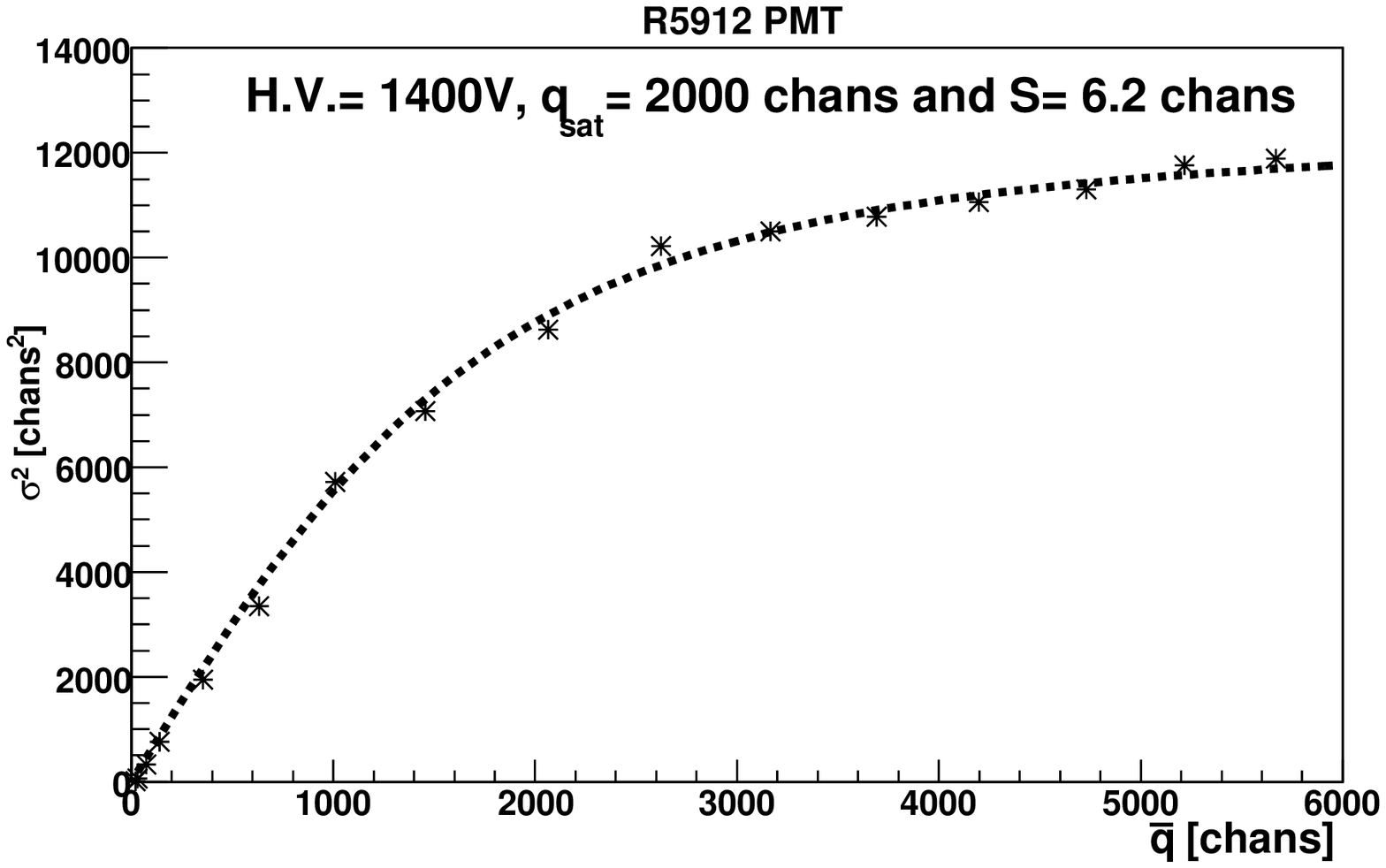}\end{center}
\begin{center}\includegraphics[%
 scale=0.43 ]{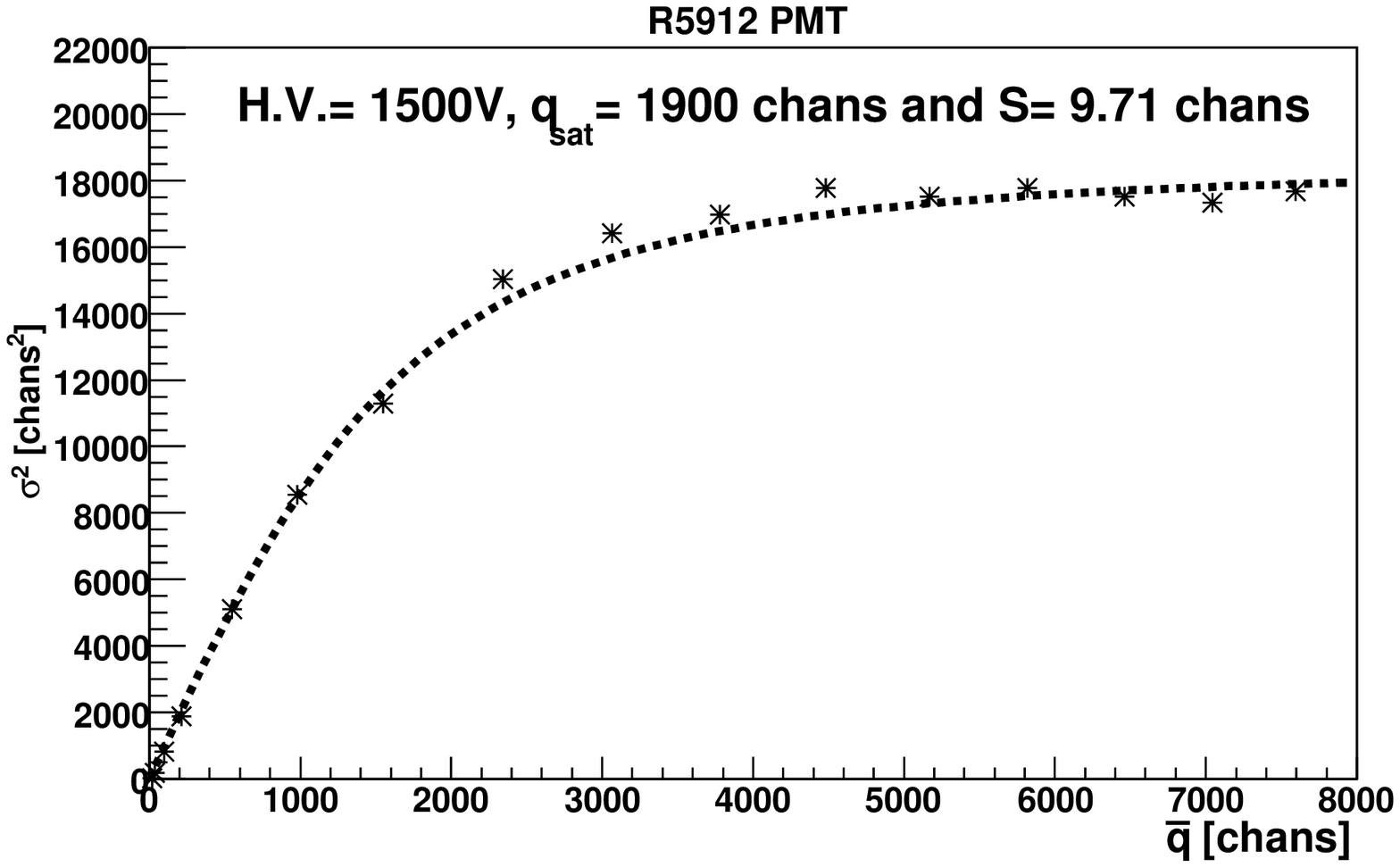}\end{center}
\begin{center}\includegraphics[%
 scale=0.43 ]{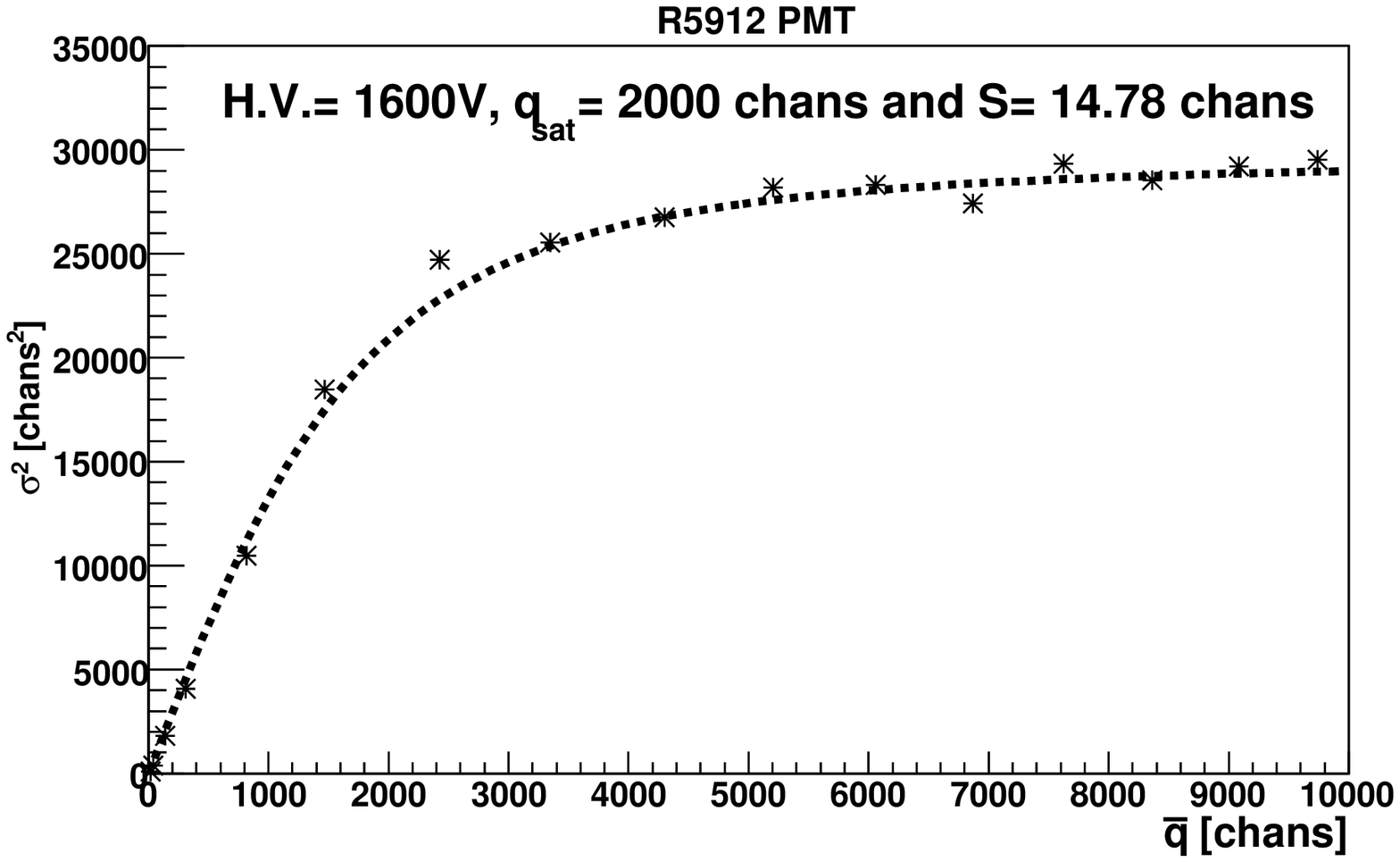}\end{center}
\caption{\label{cap:R5912_1600V} Comparison of the model with the data obtained
at different high voltages, $q_{sat}$ and S are in ADC channels.
The stars represent the data, and the dashed line represents the model with
$\alpha= 2$ for all.}
\end{figure}

$\gamma$ is a factor depending on the single photoelectron
charge distribution of the PMT. Since
the interest of this paper is to explain the saturation profile of the PMT,
it is enough to express the equation \ref{generalVar} like following:

\begin{equation}
\label{TheoGenVar}
\sigma^2=S \frac{q} {\sqrt{1+\frac{q^{\alpha}}{q_{sat}^{\alpha}}}}.
\end{equation}

where $S= \gamma e G_0$, S is the slope of the linear region when the variance
is plotted as function of the measured charge $q$.\\
Figure \ref{cap:R5912_1600V} shows the
comparison of the model with the data obtained from the R5912
PMT, this PMT came already encapsulated with its own voltage divider. The
procedure for the gain measurement is exactly the same as for the R1408 PMT
like described above. The model reproduces  the data with $\alpha=2$ in this
case, too.

\subsection{The MCP-PMT case}
In this section, I compare the model to the data obtained with a new
kind of PMT, the MCP-PMT. The ref. \cite{HAM06} contains a general
description of this type of PMTs. The data which I used
for the comparison with the model is published in the ref. \cite{Inam08}. 
To  compare to the data, the following equation is used:

\begin{equation}
\label{Gphoton1}
\frac{G}{G_0}=\sqrt{\frac{-1+\sqrt{1+8\frac{N_{\gamma}^2 }{N_{\gamma sat}^2 }}}{4\frac{N_{\gamma}^2 }{N_{\gamma sat}^2}}}.
\end{equation}

The equation \ref{Gphoton1} gives the gain variation as function of 
the number of the incident 
photons $N_{\gamma}$. $N_{\gamma sat}$ is the number of incident 
photons at which the
PMT saturation occurs. $N_{\gamma sat}$ is directly determined from the data
as the value of $N_{\gamma}$ which corresponds to the ratio 
$\frac{G}{G_0}= \frac{1}{\sqrt{2}}$. 
The demonstration of the equation \ref{Gphoton1} 
is made in the appendix. Figure \ref{cap:MCPPMT} shows the results of 
the comparison. Also in this case, the model reproduces the data.

\section{CONCLUSION}
To describe some of the  known properties of the photomultiplier tubes, a PMT
model is proposed.  The PMT
characteristics like the quantum efficiency, the collectivity efficiency and
the gain are linked together in natural way as demonstrated in
the appendix.  The model is successful
 to describe the saturation profile measured for three different PMTs
with  different single photoelectron charge distributions. The
comparisons with the measurements show that the value of the coefficient
$\alpha$ is 2. The effect of the
voltage divider - on the gain variation and the saturation performance - is
explained in the model by the variation of the ratio $\frac{q}{q_{sat}}$.

\section*{Acknowledgments}

I want express my gratitude to Prof. Charles Lane for initiating me to the
PMTs calibration techniques and for many helpful discussions. Also, I want
thank Prof. R. Svoboda for providing the R5912 PMT.
  \renewcommand{\theequation}{A-\arabic{equation}}
  \setcounter{equation}{0}  
  \section*{APPENDIX: Formulation of the PMT characteristics}  

\subsection*{\textbf{The collected Charge at the anode and the gain:}}

The collected charge at the anode can be expressed as function of the
number of the photoelectrons, which are collected at the first 
dynode $\mu$ and the gain G as following:

\begin{equation}
\label{charge}
q=\mu G e,
\end{equation}

Where $e=1.6~ 10^{-19}C$ is the electron charge.\\
Now, if the equation \ref{gain} with $\alpha= 2$ is taken in account, 
the following equation will be obtained:
\begin{equation}
\label{master}
\frac{e^2 \mu ^2}{q^2_{sat}} G^4 + G^2 - G^2_0 = 0.
\end{equation}

The equation \ref{master} can be considered as the master equation of the PMT,
which links the observables of the PMT to each other, the quantum efficiency 
via $\mu$, the collectivity efficiency represented by $q_{sat}$ and the gain G.

The equation \ref{master} has like solution the following expression:

\begin{equation}
\label{gain_mu_2}
e G = \sqrt{\frac{-1+\sqrt{1+\frac{4 e^2 G^2_0}{q^2_{sat}} \mu ^2}}{\frac{2 \mu ^2} {q^2 _{sat}}}}.
\end{equation}

 The equations \ref{charge} and \ref{gain_mu_2}  let to express 
the collected charge at the anode, like following:

\begin{equation}
\label{N_mu_2}
q=\sqrt{\frac{-1+\sqrt{1+\frac{4 e^2 G^2_0}{q^2_{sat}}\mu^2}}{\frac{2}{q^2_{sat}}}}.
\end{equation}

In the case of the cathode current saturation, $\mu$ becomes constant which 
makes the  collected charge at the anode and the gain constants, too.

\subsection*{\textbf{The quantum efficiency:}}

If the equations \ref{gain} and \ref{charge} are put together, $\mu$ can be
expressed as function of the collected charge at the anode, q, as following:

\begin{equation}
\label{mu_q}
\mu=\frac{\sqrt{q^2(1+\frac{q^2}{q^2_{sat}})}}{e G_0}.
\end{equation}

$\mu$ is linked to the number of incident photons, $N_{\gamma}$, by the following relation:
\begin{equation}
\label{mu_QE}
\mu=\rho\eta N_{\gamma},
\end{equation}

Where $\rho$ is the collectivity efficiency between the photocathode and the 
first dynode, $\eta$ is the quantum efficiency of the PMT.

 The equations \ref{mu_q} and \ref{mu_QE} let to express the product
$\rho \eta$ as following: 

\begin{equation}
\label{QE}
\rho \eta = \frac{\sqrt{q^2(1+\frac{q^2}{q^2_{sat}})}}{e G_0 N_{\gamma}}.
\end{equation}

\subsection*{\textbf{The number of  photoelectrons at the saturation:}}

For $q=q_{sat}$, the equation \ref{mu_q} gives the number of photoelectrons at
the saturation, $\mu_{sat}$, collected at the first dynode:

\begin{equation}
\label{qsat}
\mu_{sat}=\sqrt{2}~\frac{q_{sat}}{eG_0},
\end{equation}
$\mu_{sat}$ characterizes the saturation level of the PMT at a given
high voltage. If the equation \ref{G0} is used with the equation \ref{qsat},
$\mu_{sat}$ can be expressed as function of the applied high voltage:

\begin{equation}
\label{qsat_HV}
\mu_{sat}=\sqrt{2}~\frac{q_{sat}}{e k V^{\beta N}}.
\end{equation}

Since the value of $q_{sat}$ depends only on the type of the voltage divider, 
the number of the photoelectrons
at the saturation decreases with increasing applied high voltage for a given 
PMT with a given voltage divider.\\
Using the equations \ref{gain_mu_2}, \ref{mu_QE} and \ref{qsat}, 
 the gain can be written as function of the number 
of incident photons $N_{\gamma}$:
\begin{equation}
\label{Gphoton}
\frac{G}{G_0}=\sqrt{\frac{-1+\sqrt{1+8\frac{N_{\gamma}^2 }{N_{\gamma sat}^2 }}}{4\frac{N_{\gamma}^2 }{N_{\gamma sat}^2}}},
\end{equation}
Where $N_{\gamma sat}= \frac{\mu_{sat}}{\rho \eta}$.

\newpage

\begin{figure}[h]
\begin{center}\includegraphics[%
 scale=0.5 ]{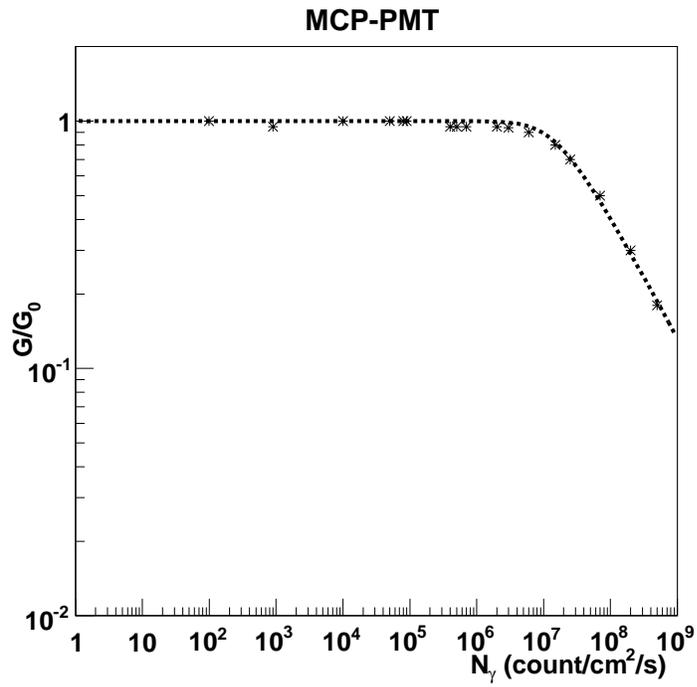}\end{center}
\caption{\label{cap:MCPPMT} Relative gain $\rm{\frac{G}{G_0}}$ vs photon rate 
$\rm{N_\gamma}$. The dashed line represents the model 
(i.e. the equation \ref{Gphoton1} with 
$\rm{N_{\gamma sat}= 2.5~10^7 count/cm^2/s}$), and the stars represent 
the data from the ref. \cite{Inam08}.}
\end{figure}

\end{document}